\documentclass[ams,preprint,showpacs]{revtex4}

  \usepackage{graphicx,natbib}
  \usepackage{bm,amsmath,amsfonts,color}

\usepackage{natbib}
\usepackage{times}

\def \beq {\begin{equation}}
\def \eeq {\end{equation}}
\def \bea {\begin{eqnarray}}
\def \eea {\end{eqnarray}}
\def \bfig {\begin{figure}}
\def \efig {\end{figure}}

\def \fig {Fig.~\ref}

\begin{document}

\title{Proton Kinetic Effects in Vlasov and Solar Wind Turbulence}

\author{S. Servidio$^{1}$, K.T. Osman$^{2}$, F. Valentini$^{1}$, D. Perrone$^{1}$, F. Califano$^{3}$, 
S. Chapman$^{2}$,  W. H. Matthaeus$^{4}$, and P. Veltri$^{1}$}
\affiliation{
$^1$Dipartimento di Fisica, Universit\`a della Calabria, I-87036 Cosenza, Italy\\
$^2$Centre for Fusion, Space and Astrophysics; University of Warwick, Coventry, CV4 7AL, United Kingdom\\
$^3$Dipartimento di Fisica and CNISM, Universit\`a di Pisa, 56127 Pisa, Italy\\
$^4$Bartol Research Institute and Department of Physics and Astronomy, University of Delaware, Newark, DE 19716, USA
} 
\date{\today}

\begin{abstract} 
Kinetic plasma processes have been investigated in the framework of solar wind turbulence, 
employing Hybrid Vlasov-Maxwell (HVM) simulations.
The dependency of proton temperature anisotropy $T_\perp/T_{\parallel}$ on 
the parallel plasma beta $\beta_{\parallel}$, commonly observed 
in spacecraft data, has been recovered 
using an ensemble of HVM simulations. By varying plasma parameters, 
such as plasma beta and fluctuation level, the simulations explore
distinct regions of the parameter space given by $T_\perp/T_{\parallel}$ and $\beta_{\parallel}$, 
similar to solar wind sub-datasets. Moreover, both simulation and solar wind data 
suggest that temperature anisotropy is not only associated with 
magnetic intermittent events, but also with 
gradient-type structures in the flow and in the density. 
This connection between non-Maxwellian kinetic effects and various types of intermittency may be a 
key point for understanding the complex nature of plasma turbulence.
\end{abstract}

\pacs{52.35.Ra, 96.50.Ci, 52.65.-y, 94.05.-a, 94.05.Lk, 52.65.Ff}

%52.35.Ra -> Plasma turbulence 
%96.50.Ci -> Solar wind plasma; sources of solar wind 
%52.65.-y -> Plasma simulation
%94.05.-a -> Space plasma physics (see also 96.50.-e Interplanetary physics)
%94.05.Lk -> Turbulence 
%52.65.Ff -> Fokker-Planck and Vlasov equation 

\maketitle

In magnetohydrodynamic turbulent flows, regions of strong gradients define small scale
coherent structures that are expected to be sites of enhanced dissipation \cite{Veltri99}. 
Such coherent structures 
may also be sites of magnetic reconnection and plasma heating 
\cite{marsch1parker}. 
On the other hand, in low-collisionality plasmas, such as the 
solar wind,  it is expected that 
kinetic processes lead to other phenomena such as  
temperature anisotropy and energization of suprathermal 
particles \cite{marsch1parker,Hellingergary,OsmanEAL}.
Since solar wind plasma is generally observed to be in a 
strongly turbulent state it is far from clear how processes 
such as dissipation operate, and how 
observed microscopic non-equilibrium conditions are related to 
the dynamics and thermodynamics that influences the large scale features, 
including the origin and acceleration of the solar wind itself.

Far from the textbook conditions of uniform plasma equilibrium, 
that motivate much of the traditional discourse on plasma dissipation, 
the highly excited but weakly collisional solar wind 
demonstrates  a more complex relationship between 
the macroscopic state and the microscopic physics than one 
would find in a viscous fluid. 
Here we explore the connections between 
turbulence and solar wind properties, 
employing Vlasov kinetic simulations.
We find that the simulations are able to recover solar wind kinetic 
phenomena through the combined effect of reasonable variation in the 
initial parameters along with the natural dynamical variations 
produced by the turbulence itself. Therefore, we suggest that 
the kinetic properties of an ensemble of solar wind observations
is controlled by turbulence properties.

In situ spacecraft measurements reveal that interplanetary 
proton velocity distribution functions (VDFs) are anisotropic with respect to the 
magnetic field \cite{marsch82tu04}.
Values of the anisotropy $T_\perp/T_{\parallel}$  range broadly, with most values 
between $10^{-1}$ and $10$ \cite{BaleEA09,MarucaEA11}.
The distribution of $T_\perp/T_{\parallel}$ depends systematically 
on the ambient proton parallel beta $\beta_{\parallel}=n_p k_B T_{\parallel}/(B^2/2\mu_0)$ -- the ratio of 
parallel kinetic pressure to magnetic pressure, manifesting a characteristic shape 
in the parameters plane defined 
by $T_\perp/T_{\parallel}$ and 
$\beta_{\parallel}$ \cite{Gary,BaleEA09,MarucaEA11}. 
More recently \cite{OsmanEA12}, observations have suggested 
that a link exists between anisotropy and intermittent current sheets.
The latter study employed 
the Partial Variance of Increments (PVI) technique
which provides a running measure of the 
magnetic field intermittency level,
and is able to quantify the presence of strong 
discontinuities \cite{GrecoEA08}.  Elevated PVI values signal an increased 
likelihood of finding coherent magnetic 
structures such as current sheets, and occur in the same regions of parameter space 
where elevated temperatures are found \cite{OsmanEA12}, 
and also near to identified instability thresholds \cite{MarucaEA11}. 
Hybrid-Vlasov and Particle In Cell simulations
of turbulence complement  these findings by establishing 
that kinetic effects are concentrated near  regions of strong 
magnetic stress \cite{ServidioEA12,drakeEAL,PerroneEA13,WanEA12,KarimabadiEA13}. 
Here we further investigate 
this path by exploring a broad range of plasma parameters, 
and establishing a more complex link between temperature anisotropy and turbulence intermittency.

\begin{figure}  %%[hbt]
\begin{center}
\includegraphics[width=0.8\columnwidth]{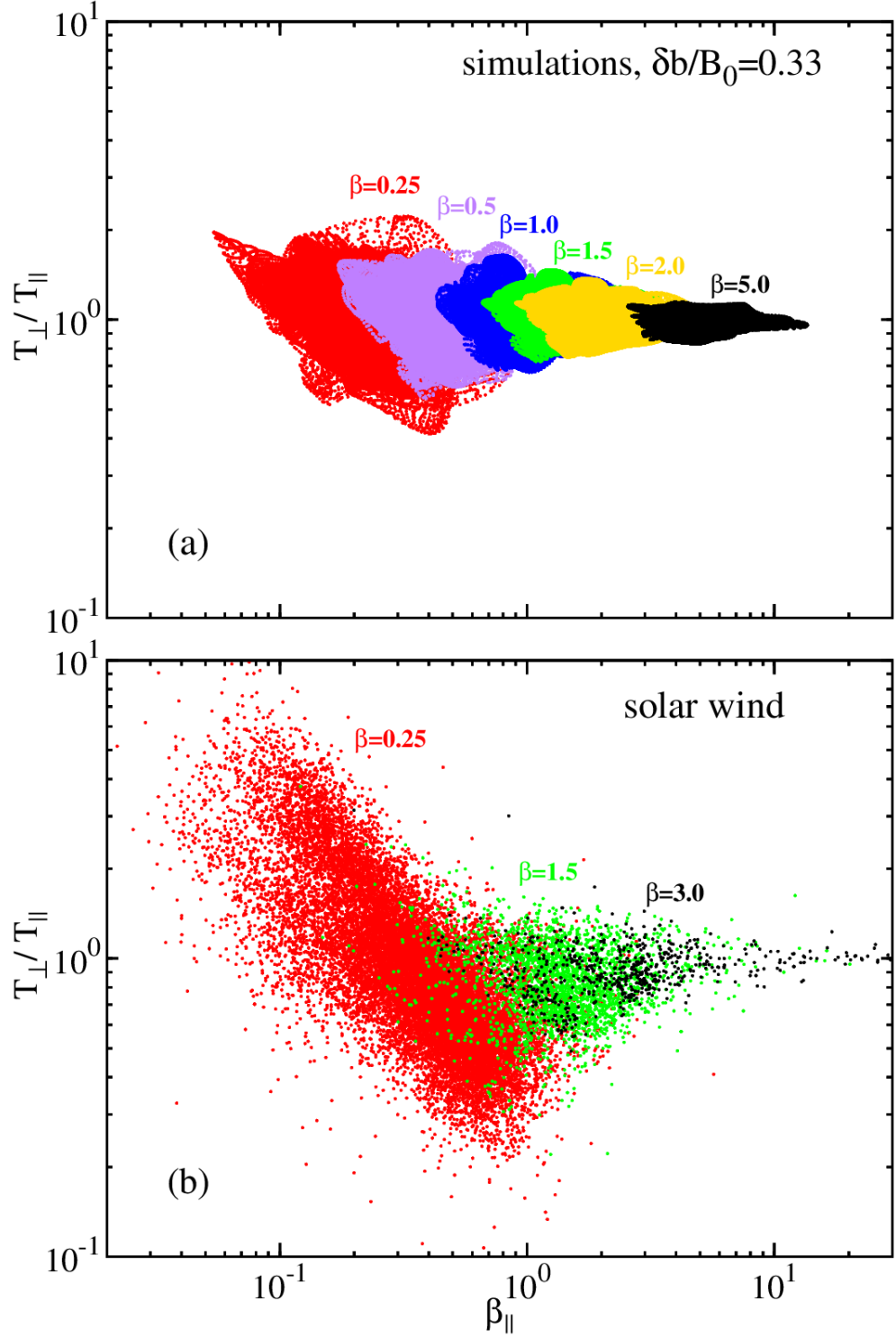}
\end{center}
\caption{ (Color online) (a) Scatter plot of anisotropy $T_\perp/T_{\parallel}$
{\it vs.} $\beta_\parallel$ for the HVM simulations, performed
with $\delta b/B_0=0.33$, and varying $\beta=0.25,0.5,1.0,1.5,2.0$ 
and $5.0$ (from left to right). 
(b) Solar wind samples in the same plane, 
sorted in four hour samples with the average value of $\beta=0.25$ (red), 1 (green), and 3 (black).}
\label{fig:scat}
\end{figure}

Kinetic plasma turbulence is an incompletely understood problem, 
and treatments such as linear and quasi-linear simplifications of the Vlasov-Maxwell 
equations may provide useful guidance \cite{davidson}.
However, for plasmas found to be in a turbulent state, 
it is not at all obvious whether such simplified models reliably  provide a valid description. 
On both technical and physical grounds, one might
question whether linear homogeneous 
Vlasov theory is sufficient to explain the inhomogeneous plasma 
dynamics operating near coherent structures.
Hence, a strong basis for 
analyzing the dynamics of such plasmas 
is provided by direct numerical simulations of plasma kinetic equations, 
in which the time evolution of the VDF is 
described self-consistently, and in the absence of particle noise (a crucial point in 
studying small scale gradients \cite{HaynesEA13}.) 
In turbulent systems such as the 
solar wind \cite{carboneLivRev}, it is of crucial relevance to quantify 
the role of kinetic effects in the turbulent cascade, since this provides a path 
to explain the energy dissipation mechanisms. 
Non-Maxwellian features of the VDF represent a direct manifestation of the underlying 
complex kinetic processes. Here we perform an ensemble of direct numerical simulations of the 
Hybrid Vlasov-Maxwell (HVM) model \cite{ValentiniEAL1}. We compare results with solar wind 
datasets from the Wind spacecraft, 
and we investigate the structures that contribute to the local anisotropy 
observed in the solar wind.

We performed direct numerical simulations of a 5 dimensional 
(2D in space; 3D in velocity space) 
Vlasov model \cite{ValentiniEAL1,ServidioEA12} for protons, coupled
to a fluid model for electrons. 
The 2D plane is perpendicular to the mean field $B_0\hat{z}$, 
and fluctuating vectors are 3D. 
In order to mimic the variability of the solar wind, we vary the plasma beta, 
and also the level of fluctuations $\delta b/B_0$, 
where $\delta b$ the rms fluctuation value. 
The simulation box is of size $2\pi 20 d_i$ ($d_i$ is the ion skin depth), 
with a resolution of $512\times 512$ in the physical space, 
and a typical resolution of $51^3$ in the velocity space.
The velocity space resolution is varied for 
the simulations with smaller plasma beta, 
where we tested the results by varying the resolution from $51^3$ to $81^3$.
For these parameters, the conservation of the total mass and energy of the system in
the simulations is satisfied with typical relative errors of 
$\simeq 10^{-3}\%$ and $\simeq 10^{-5}\%$, respectively.
%As described in \cite{ServidioEA12}, to initialize the turbulence, 
%we specify an initial (Gaussian) spectrum of 
%fluctuations, and an isotropic Maxwellian plasma 
%($T_\perp/T_{\parallel}=1$) with uniform temperature.  
%The large scale bend-over in the spectrum 
%fixes the correlation length to be $\simeq 10 d_i$. 
%The range of scales 
%included in the  simulation is a 
%compromise between 
%a reasonable representation of small scale effects, 
%emphasizing proton kinetic effects,  
%and a description that extends into the fluid regime.
%Such a 
%balance is important in a statistical study of turbulence, 
%in which both small scale effects and larger scale 
%statistical homogeneity need to be taken into account. 
%(Resolving a correlation length, even if not realistically separated from
%kinetic scales, makes it possible to have convergent statistics, e.g., 
%in defining a mean field $B_0$.)
As described in \cite{ServidioEA12},
we initialize the turbulence
by specifying a band limited Gaussian
spectrum of fluctuations,
and an isotropic Maxwellian plasma ($T_\perp/T_{\parallel}=1$) with uniform temperature.
The correlation length (energy containing scale)
is $\ell \simeq 10 d_i$.
The range of dynamically accessible
scales is a compromise due to a finite simulation
size, but it includes
both proton kinetic scales and
extends into the fluid regime.
This class of simulations evolves
\cite{ServidioEA12,GrecoEA12}
by forming a broad band spectrum extending
from correlation scale $\ell$
to kinetic scales ($<d_i$),
implying an effective Reynolds number, as in
classical turbulence theory,
on the order of $(\ell/d_i)^{4/3}$,
while also forming characteristic
small scale structures associated with
intermittency. Therefore the dynamics appears to be analogous
to moderately high ($\gg1$) Reynolds number strong
turbulence. 
%Moreover, it is worth noting that solving this range of scales 
%is important in a statistical study of turbulence, 
%in which both small scale effects and larger scale 
%statistical homogeneity need to be taken into account. 

For each simulation we used the data 
near the time of 
peak of nonlinear activity \cite{ServidioEA12}. 
A scatter plot of temperature anisotropy 
as a function of the $\beta_{\parallel}$
is shown in \fig{fig:scat}, 
for  simulations initialized 
with $\delta b/B_0=1/3$, and with uniform initial 
plasma beta varying over 
values $\beta=0.25,0.5,1,1.5,2,5$.
It is apparent that the dynamically evolved data 
are strongly modulated by the choice of beta, and are also 
spread in temperature anisotropy (note that at $t=0$, $T_\perp/T_{\parallel}=1$).
Notably, the resulting distributions
resemble the familiar form of those  accumulated from years of
solar wind data, as in \cite{BaleEA09,OsmanEA12}.

In order to further confirm our methodology, 
a similar analysis has been carried out using a large sample 
of solar wind data,  
binning the data according to 
plasma $\beta$. 
The solar wind dataset, which spans 17 years with a cadence of 
92 seconds,  
is divided into 4-hour non-overlapping datasets (about 5 correlation lengths).
These are sorted into three bins having average values 
of $\beta=0.25\pm0.01, 1.5\pm0.01, 3\pm0.01$, 
where all the data falling outside of this range is excluded.
As can be seen from \fig{fig:scat}-(b), 
when the data 
are sorted according to their average $\beta$ in this way, 
the patterns of data
in the plane 
move from left to right in the plot,
spanning systematically the plane, in good 
agreement with the simulations.
Since we know that in the simulations 
non-Maxwellian  kinetic effects such as temperature
anisotropies are concentrated in the non-Gaussian 
coherent structures \cite{ServidioEA12}, 
the above result 
confirms the major role that 
kinetic turbulence plays in the macroscopic 
distribution of non-Maxwellian effects in the solar wind.

\begin{figure}  %%[hbt]
  \begin{center}
    \includegraphics[width=0.8\columnwidth]{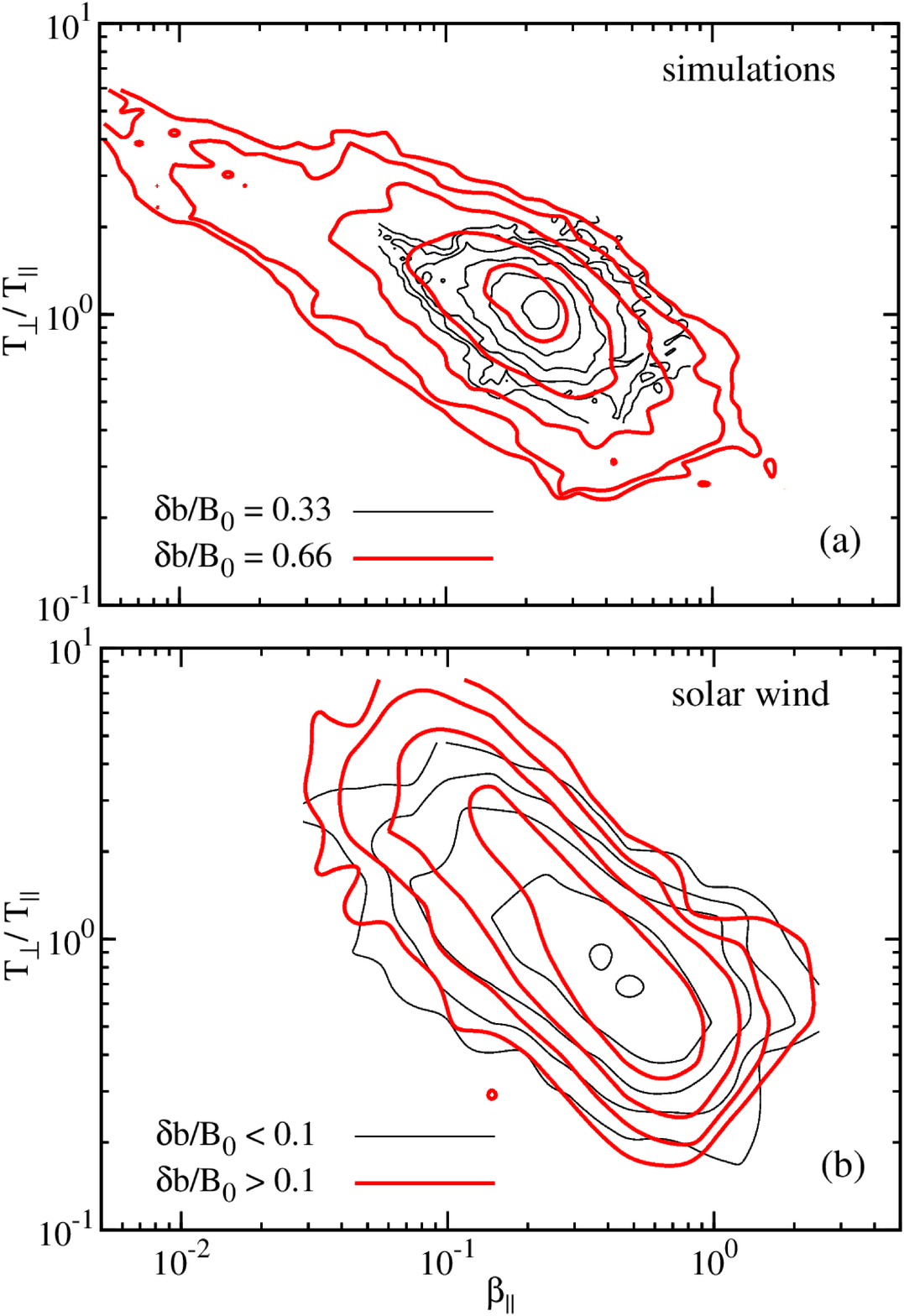}
  \end{center}
  \caption{
(Color online) (a) Joint distributions of 
$T_\perp/T_{\parallel}$ {\it vs.} $\beta_{\parallel}$, 
comparing simulations with $(\beta,\delta b/B_0)=(0.25,1/3)$ (thin-black) and 
$(0.25,2/3)$ (thick-red). (b) Samples of solar wind selected for four hour average 
values of $\beta=0.25$, and $\delta b/B_0<0.1$ (thin-black), and $>0.1$ (thick-red).
}
    \label{fig:compdbb0}
\end{figure}

%In order to further confirm our methodology to study the temperature anisotropy in space plasmas, 
%a similar analysis has been carried out using a large sample 
%of solar wind, 
%binning the solar wind by plasma $\beta$. 
%The solar wind dataset, which spans 17 years, 
%is divided into 4-hour nonoverlapping datasets (about 5 correlation lengths).
%These are sorted according to their average plasma $\beta$.
%Naturally  there is some variability in the values of $\beta$ and anisotropy 
%with in each sample.  In the bottom panel of \fig{fig:scat}, 
%the solar wind data of anisotropy 
%versus the ambient $\beta_{\parallel}$, are reported for several values of $\beta$. 
%When these XXX (KAREEM) second samples are plotted in the $\beta_\parallel, R$ plane, 
%this variability causes a substantial spread of the points.
%When the samples of solar wind data are color coded according to the
%average values of $\beta$ in the sample from which they are obtained,
%the pattern that appears (Fig.\ref{fig:scat})
%can be seen to be in good agreement with the 
%general  shape of the distributions seen in the simulations.
%Since we know that in the simulations  
%non-Maxwellian  kinetic effects such as temperature
%anisotropies are concentrated in the non-Gaussian
%coherent structures, 
%the above result 
%confirms the major role that kinetic turbulence plays in the macroscopic 
%distribution of non-Maxwellian effects in the solar wind. 

The distribution determined from simulations 
shows the apparent signatures of 
regulation of the anisotropy 
frequently
associated with instability thresholds, 
even though the envelope of the distribution 
appears to be 
somewhat further away from the reported  
mirror and firehose instability thresholds
in the solar wind analyses \cite{BaleEA09,MarucaEA11}. 
This suggests 
that the level of turbulent fluctuations, here represented 
by $\delta b/B_0$, may play another 
important role in the explanation of the observed anisotropy. 
To examine 
the influence of turbulence level
on these 
distributions, 
we performed a set of simulations varying the level 
of fluctuations from $\delta b/B_0=1/3$ to $2/3$. 
In \fig{fig:compdbb0} we compare 
data density (PDFs) of simulations with $(\beta,\delta b/B_0)=(0.25,1/3)$ and $(0.25,2/3)$.
It is evident that the level of 
fluctuations, together with the mean plasma beta, strongly 
influences the distribution of anisotropies in Vlasov turbulence. 
Similar results have been obtained for the case with 
$(\beta,\delta b/B_0)=(1,1/3)$ and $(1,2/3)$ (not shown in this plot).  
A similar analysis has been carried out for the solar wind, 
sampling the data for both $\beta$ and $\delta b/B_0$;
see \fig{fig:compdbb0} (bottom).  This further confirms, especially 
for low beta, that the level of fluctuations plays a major role in 
shaping the distribution  of temperature anisotropies.

A consistent interpretation of the above results is that the turbulent dynamics produces
variations in kinetic anisotropies (measured here by $T_\perp/T_{\parallel}$ and $\beta_\parallel$)
even when the global average values are prescribed. 
Furthermore, when the global average values of $\beta$ 
and $\delta b/B_0$ are varied, 
the dynamical spreading of 
local anisotropies ventures into different, and sometimes 
more distant regions of the parameter space. 
This effect is observed to be qualitatively
similar in the simulations and in the solar wind analysis, 
keeping in mind of course that the control 
over parameters is direct in the former case, and 
obtained through conditional sampling in the latter.
This interpretation may be expanded further, 
taking into account recent studies that show 
concentrations of kinetic effects near coherent 
structures. These effects include
elevated temperatures, and enhanced kinetic
anisotropies, and are seen in plasma simulations
\cite{ServidioEA12,WanEA12,KarimabadiEA13}
and in solar wind observations \cite{OsmanEAL,OsmanEA12,OsmanEA12b}.
One might reason in this way: intermittency is a generic feature 
of turbulence, leading to coherent structures of increasing 
sharpness at smaller scales, the effect growing stronger at higher
 Reynolds numbers \cite{AntoniaSreenivasvan}.
Stronger fluctuation amplitude is associated with stronger 
turbulence (e.g., higher Reynolds number, larger cascade rate), and therefore
for a plasma, larger $\delta b/B_0$ should be associated with 
stronger intermittency and stronger small scale coherent structures.
Since coherent structures are connected with kinetic anisotropies, then
larger $\delta b/B_0$ should also be connected with 
stronger anisotropies.
A natural explanation for the current observations emerges from this 
line of reasoning.
 It is noteworthy that this  is an alternative
to the interpretation put forth previously \cite{BaleEA09}
 that the fluctuation levels are larger near the 
parameter space regions having larger anisotropies because 
instabilities that operate in those regions 
also act to excite these fluctuations.
This interpretation 
provides an alternative
in which the anisotropies are a consequence of the turbulence.

\begin{figure}  %%[hbt]
  \begin{center}
    \includegraphics[width=0.8\columnwidth]{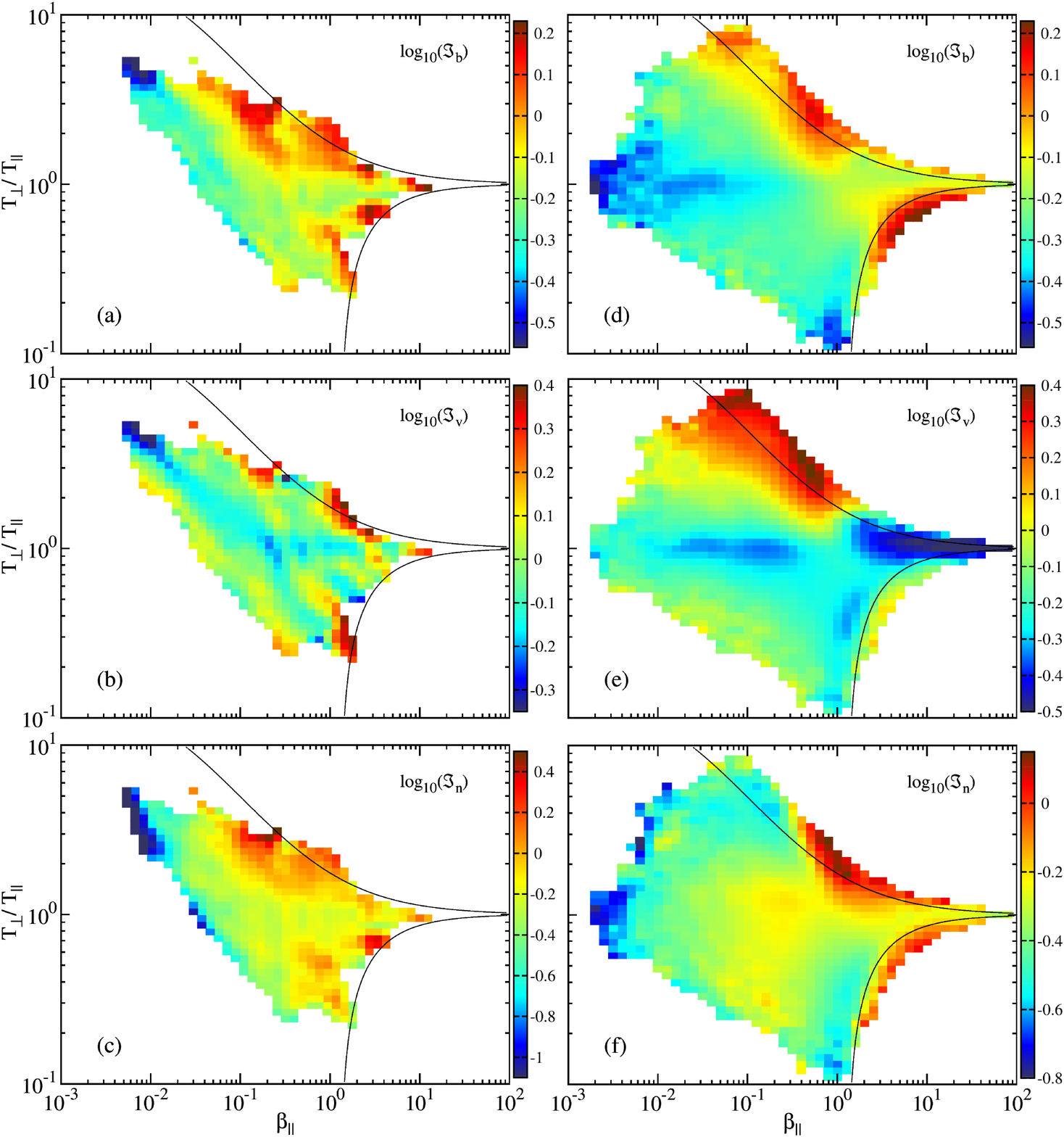} 
  \end{center}
  \caption{(Color online) Average $\Im_f$ in the anisotropy-$\beta_\parallel$ plane for the ensemble of simulations: $\Im_b$ (a), $\Im_v$ (b) and $\Im_n$ (c). Same for the solar wind, in the panels (d)-(e). In each panel, dashed curves indicate 
theoretical growth rates for the mirror ($T_\perp/T_{\parallel}>1$) and 
the oblique firehose ($T_\perp/T_{\parallel}<1$) instability.
}
    \label{fig:allbet1}
\end{figure}

At this point, since from \fig{fig:compdbb0}-(top) it is evident that 
each simulation has different boundaries in the anisotropy plane, 
it is instructive 
to ask whether there is an association 
between structures and the observed anisotropy. 
Such a connection was already established in the solar wind
\cite{OsmanEA12} based on analysis of magnetic fluctuations. 
Therefore 
recent discussion has focused on 
intermittency of the 
magnetic field 
and its connection to 
the observed anisotropy.
In plasma turbulence, however, 
dynamical couplings may also involve structure in 
other fields, 
and other candidates for the association with 
anisotropy cannot be excluded.
Here we employ both simulations and solar wind data to explore this possibility, 
analyzing the association of magnetic, density and velocity gradients 
with the occurrence of strong kinetic effects. 

In analogy with previous work on 
magnetic intermittency \cite{GrecoEA08,GrecoEA12}, 
here we employ a PVI analysis
for examination  of flow and density gradients.
This intermittency measure
is given by 
\beq
\Im_f(s)=\frac{|\Delta {\bf f}| }{\sqrt{\langle|\Delta {\bf f}|^2}\rangle},~\text{where}~\Delta {\bf f}={\bf f}(s+\Delta s)-{\bf f}(s),
\label{eq:pvi}
\eeq
where 
${\bf f}$ can be the magnetic (${\bf b}$) 
or velocity (${\bf v}$) vector field, 
or the scalar density field (n).
The brackets $\langle \dots \rangle$ denote
an appropriate time average over many correlation times (the entire simulation box, or the entire solar wind dataset). 
For the simulations, the variable $s$ is 
a 1D variable that spans all of the simulation 
domain, while in the solar wind, it labels
the time series at the spacecraft.

Once the data  has been 
binned in the ($\beta_\parallel, T_\perp/T_{\parallel}$) plane,
we evaluated the average magnitude of $\Im_f$ in each bin, 
using all the HVM simulations
presented in the present work. 
As can be seen 
from \fig{fig:allbet1}-(a), where $\Im_b$ is shown, the strongest magnetic gradients
are found near the threshold regions, 
in agreement with \cite{OsmanEA12,ServidioEA12}. 
These can be current sheets or other discontinuities.
In panel (d) of the same figure, the same
analysis is shown for 
17 years of solar wind data, 
indicating that magnetic gradients are 
likely playing a role in the observed anisotropy. 
We also performed the 
same analysis for the velocity field, obtaining 
$\Im_v$, which is a surrogate for the 
vorticity of the flow.
It can be seen in \fig{fig:allbet1}-(b)
that intermittency of the velocity is also strongly  
correlated to kinetic anisotropy.
The signatures are once again 
near the boundaries of the
characteristic anisotropy plot. 
Finally, the bottom panels of \fig{fig:allbet1} report
the same analyses for $\Im_n$, and the  
results qualitatively resemble 
the magnetic and velocity field cases. 
It is noteworthy that 
the PVI analyses of velocity and density 
for the solar wind cases, 
\fig{fig:allbet1} (d)-(e),
also show interesting features near the boundaries. 
Note that apart from the reasonable agreement 
in the shape of the distributions, 
the values of the $\Im$ field are 
comparable between the simulations and solar wind data. 
Note also that the simulations 
show a more discontinuous behavior of the PVI very probably due to the
discretization of the simulation. This effect will eventually disappear for
a much higher number of simulations, that would finally realize the ergodicity of the solar wind.

These analyses converge essentially toward the same conclusions, namely that
in plasma turbulence there is a strong link between intermittent structures  and kinetic anisotropy.
The multiple analyses presented here suggests that the intermittent structures, 
both magnetic and kinetic, may be central ingredients 
in sustaining the observed
anisotropy.
For example, structures may locally be found in near-equilibrium conditions, and 
in the absence of collisions, 
such configurations 
might require 
a certain amount of temperature anisotropy \cite{mikha}.

The present study demonstrates further that
it is possible to extract statistical features from 
kinetic plasma simulations that motivate 
interpretations of solar wind behavior based 
on fully nonlinear plasma physics.
Such studies may be carried out without 
recourse to extreme assumptions such as uniform 
plasma equilibria or linear Vlasov waves and instabilities. 
Here we employed nonlinear Hybrid-Vlasov simulations
in 2.5 dimensions to show that:
(1)
the initially controlled average plasma beta and fluctuation level
produces turbulent dynamics that leads to a 
pointwise spread in the ($\beta_{\parallel}$, $T_{\perp}/T_{\parallel}$)
plane, reminiscent of solar wind populations in the same parameter
plane; (2) simulations with moderate variations of average
$\beta$ and $\delta b/B_0$ lead to fuller coverage of the 
$(\beta_\parallel,T_\perp/T_{\parallel})$ plane, a tendency that is
reproduced by conditional sampling of a large number
of solar wind datasets; (3) 
that the simulations naturally lead to stronger $\delta b/B_0$ near 
the boundaries of the distribution;  and (4) that the 
extreme regions of the distribution of points in  
$(\beta_\parallel, T_\perp/T_{\parallel})$ plane also show enhanced values of
magnetic field gradients,
velocity shears, and density gradients.
All of these features, corroborated here by observations, point to an interpretation 
in which the appearance of large kinetic anisotropies is connected
to intermittent turbulence, and the dynamical appearance of 
coherent structures where intense anisotropy is produced. 

In the present analysis we have not been able to examine additional 
effects that may also be important in controlling kinetic anisotropies.
For example, three dimensional effects may contribute in significant 
ways.  
The kinetic response of electrons, not explored here,
may be interesting as well
and has been recently implicated in producing coherent structures
(see e.g., \cite{KarimabadiEA13}).
Finally, it is known that expansion produces 
important and systematic effects in the solar wind
that have a major impact on the evolution of kinetic anisotropy;
see e.g., \cite{HellingerEA}. 
Further and more elaborate simulations and analysis will be required to
incorporate all of these effects in a single study. However, 
we suspect that 
greater realism will show additional effects while the basic features 
we have described will persist: 
intermittent turbulence and coherent structures
have significant influence on the development of 
kinetic effects in a low collisionality plasma such as the solar wind. 

This research was partially supported by the Turboplasmas project 
(Marie Curie FP7 PIRSES-2010-269297), POR Calabria FSE
2007/2013, the NASA 
Magnetosphere Multiscale Mission Theory and Modeling program
NNX08AT76G, UK STFC, the Solar Probe Plus ISIS project, the NSF SHINE (AGS-1156094) and 
Solar Terrestrial (AGS-1063439) Programs, 
MIUR (PRIN 2009, 20092YP7EY). Numerical simulations 
were performed on the FERMI supercomputer at CINECA (Bologna, Italy)
within the European project PRACE Pra04-771.


\begin{thebibliography}{99}

\bibitem{Veltri99}
P. Veltri, Plasma Phys. Control. Fusion {\bf 41} A787 (1999). 

\bibitem{marsch1parker}
E. Marsch, Living Rev. Solar Phys. {\bf 3}, 1 (2006); E. N. Parker, Astrophys. J. {\bf 330}, 474 (1988).

\bibitem{Hellingergary}
S. P. Gary, {\it Theory of Space Plasma Microinstabilities} (Cambridge University Press, Cambridge, 1993); 
P. Hellinger {\it et al.}, Geophys. Res. Lett. {\bf 33}, L09101 (2006).

\bibitem{OsmanEAL}
K. T. Osman {\it et al.}, Astrophys. J. Lett. {\bf 727}, L11 (2011).

\bibitem{marsch82tu04} 
E. Marsch {\it et al.}, J. Geophys. Res. {\bf 87}, 52 (1982); 
C.-Y. Tu, E. Marsch, and Z.-R. Qin,  J. Geophys. Res. {\bf 109}, A05101 (2004).

\bibitem{BaleEA09}
S. D. Bale {\it et al.}, Phys. Rev. Lett. {\bf 103}, 211101 (2009).

\bibitem{MarucaEA11}
B. A. Maruca, J. C. Kasper, and S. D. Bale, Phys. Rev. Lett. {\bf 107}, 201101 (2011).

\bibitem{Gary}
J. C. Kasper, A. J. Lazarus, and S. P. Gary, Geophys. Res. Lett. {\bf 29}, 20 (2002).

\bibitem{OsmanEA12}
K. T. Osman {\it et al.}, Phys. Rev. Lett.  {\bf 108}, 261103 (2012).

\bibitem{GrecoEA08}
A. Greco {\it et al.}, Geophys. Res. Lett. {\bf 35}, L19111 (2008).

\bibitem{ServidioEA12}
S. Servidio {\it et al.}, Phys. Rev. Lett. {\bf 108}, 045001 (2012).

\bibitem{drakeEAL}
J. F. Drake {\it et al.}, Astrophys J. {\bf 709}, 963 (2010).

\bibitem{PerroneEA13}
D. Perrone {\it et al.}, Astrophys. J. {\bf 762}, 99 (2013).

\bibitem{WanEA12}
M. Wan {\it et al.}, Phys. Rev. Lett. {\bf 109}, 195001 (2012); P. Wu {\it et al.}, Astrophys. J. Lett. {\bf 763}, L30 (2013).

\bibitem{KarimabadiEA13}
H. Karimabadi {\it et al.}, Phys Plasmas {\bf 20},  012303 (2013).

\bibitem{davidson}
R. C. Davidson, {\it Physics of Nonneutral Plasmas} (Addison-Wesley, Redwood City, CA, 1990).

\bibitem{HaynesEA13}
C. T. Haynes, D. Burgess, and E. Camporeale, arXiv:1304.1444 (2013).

\bibitem{carboneLivRev}
R. Bruno and V. Carbone, Living Rev. Solar Phys. {\bf 2}, 4 (2005); F. Sahraoui {\it et al.}, Phys. Rev. Lett. {\bf 102}, 231102 (2009).

\bibitem{ValentiniEAL1}
F. Valentini {\it et al.}, J. Comput. Phys. {\bf 225}, 753 (2007).

\bibitem{GrecoEA12}
A. Greco {\it et al.}, Phys. Rev. E {\bf 86}, 066405 (2012).

\bibitem{OsmanEA12b}
K. T. Osman {\it et al.}, Phys. Rev. Lett. {\bf 108}, 261102 (2012).

\bibitem{AntoniaSreenivasvan}
K. R. Sreenivasan and R. A. Antonia, Ann. Rev. Fluid Mech. {\bf 29}, 435 (1997). 

\bibitem{mikha}
A. B. Mikhailovskii, {\it Theory of plasma instabilities, Vol. 2: Instabilities in an inhomogeneous plasma} (Plenum, New York, 1974).

\bibitem{HellingerEA}
P. Hellinger and P. M. Tr\'avn\'icek, J. Geophys. Res {\bf 113}, A10109 (2008); L. Matteini {\it et al.}, Space Sci. Rev. {\bf 172}, 373 (2012).

\end{thebibliography}
\end{document}